\documentclass[%
 aip,
 jmp,%
 amsmath,amssymb,
 reprint,%
]{revtex4-1}

\usepackage{graphicx}
\usepackage{dcolumn}
\usepackage{bm}

\begin{document}


\title{Visualizing the melting processes in ultrashort intense laser triggered gold mesh with high energy electron radiography}
\author{Zheng Zhou}
 \affiliation{Department of Engineering Physics, Tsinghua University, Beijing 100084, China}
\author{Yu Fang}
 \affiliation{Department of Engineering Physics, Tsinghua University, Beijing 100084, China}
\author{Han Chen}
 \affiliation{Department of Engineering Physics, Tsinghua University, Beijing 100084, China}
\author{Yipeng Wu}
 \affiliation{Department of Engineering Physics, Tsinghua University, Beijing 100084, China}
\author{Yingchao Du}
 \email{dych@mail.tsinghua.edu.cn}
\affiliation{Department of Engineering Physics, Tsinghua University, Beijing 100084, China}
\author{Zimin Zhang}
\affiliation{Institute of Modern Physics, Chinese Academy of Sciences, Lanzhou 730000, China}
\author{Yongtao Zhao}
\affiliation{Xi’an Jiaotong University, Xi’an 710049, China}
\affiliation{Institute of Modern Physics, Chinese Academy of Sciences, Lanzhou 730000, China}
\author{Ming Li}
\affiliation{Institute of Applied Electronics, CAEP, Mianyang 621900, China}
\author{Chuanxiang Tang}
\affiliation{Department of Engineering Physics, Tsinghua University, Beijing 100084, China}
\author{Wenhui Huang}
\affiliation{Department of Engineering Physics, Tsinghua University, Beijing 100084, China}

\date{\today}

\begin{abstract}
High energy electron radiography (HEER) is a promising tool for high energy density physics diagnostics, apart from other tools like X/$\gamma$ ray shadowgraphy and high energy proton radiography. Impressive progresses have been made in development and application of HEER in past few years, and proved its potentials for high-resolution imaging of static opaque objects. By taking advantages of short pulse duration and tunable time structure of high energy electron probes, time-resolved imaging measurement of high energy density gold irradiated by ultrashort intense lasers has been performed. Phenomena of different time periods from picosecond to microsecond have been observed, thus proving feasibilities of this technique for imaging of static and dynamic objects.
\end{abstract}
\maketitle

\section{INTRODUCTION}

High energy density physics (HEDP) is the study of matters under extreme conditions, often designated by $>$ 1 Mbar (100 GPa) pressure or $>$ 100 GJ/m$^3$ energy density\cite{drake2006high}, such as those occurring in heavy ion\cite{Hofmann2018mre}, laser driven fusion\cite{Campbell2017mre} and similar processes. Under these conditions, hydrodynamic response of the matter is a high expansion velocity in the range of $\mu$m/ns, thus posing various challenges on diagnostic tools, such as high spatiotemporal resolution, high areal density resolution and a large dynamic range\cite{gai2014electron}. Charged particle radiography\cite{morris2013charged} has been developed as a diagnostic tool for many years by Los Alamos National Laboratory\cite{morris2006proton, morris2011flash,king1999800} and other institutes\cite{kolesnikov2010application, antipov2010radiographic,merrill2009proton}. As an alternative option to intensively studied proton radiography, high energy electron radiography (HEER)\cite{merrill2007electron} has drawn considerable interest due to its potentials to provide high spatiotemporal resolution with much easier accessibility and manipulability. Recent works have improved spatial resolution of HEER to a few microns
\cite{zhou2018experiments, merrill2018demonstration} and applied this technique to image dynamic processes\cite{merrill2018demonstration}. However, full advantages of  high energy electron probes with short pulse duration and flexibly tunable time structure have not been taken yet, and this is especially true when using high brightness electron probes generated from state-of-the-art RF photo-injectors. In a common photo-injector, pulse duration of an electron bunch can be down to ps or even fs level, and synchronization between the RF and laser can be controlled to sub-ps level. Therefore, the accuracy of time-resolved imaging of dynamic process in pump-probe scheme can reach ps level, at which level some ultrafast phenomena can be observed.

In this paper, we demonstrate the use of high brightness electron probes and a compact imaging lens composed of high gradient permanent magnet quadrupoles (PMQ) to capture the entire dynamic process of laser ablation of gold mesh over a large time scale, with tens of ps-$\mu$m spatiotemporal resolution. Successful implement of dynamic high energy electron radiograph, in ultra-fast intense laser pump, ps long high energy electron probe scheme, reveals its potential for direct visualization of fast dynamic phenomena in high energy density matters.

\section{DYNAMIC HEER EXPERIMENT}
\subsection{experimental setup}
\begin{figure*}[htb]
\includegraphics[width=0.9\textwidth]{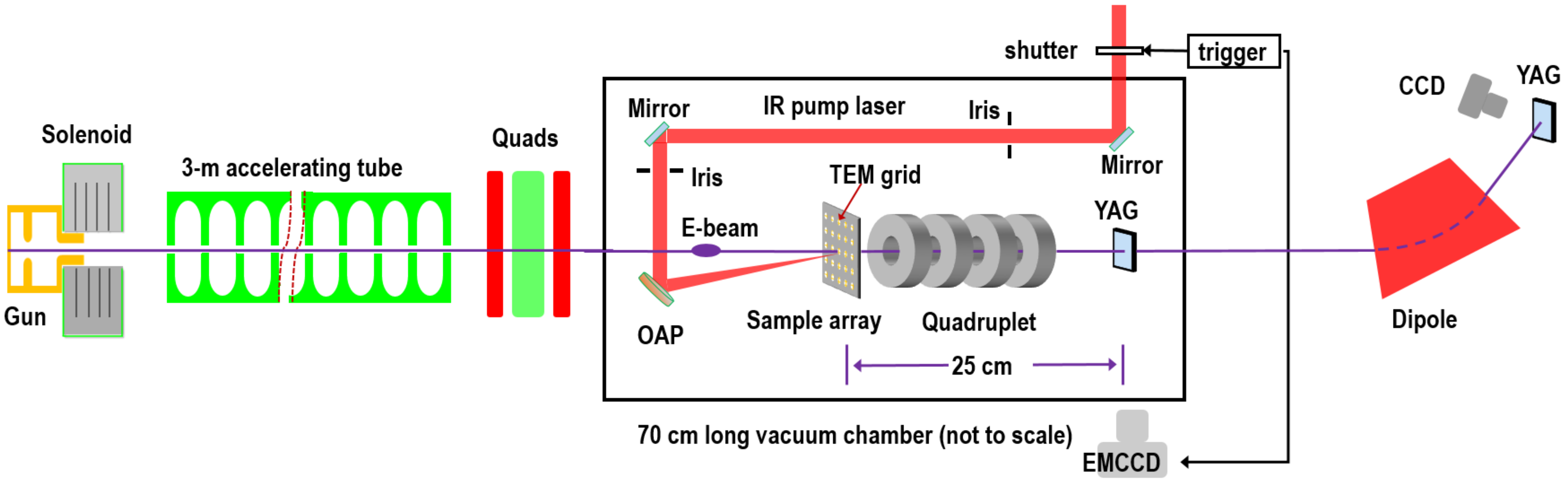}
\caption{\label{fig:1} Schematic of the dynamic HEER experimental layout. High brightness electron probes passing through the sample form a point-to-point, magnified image of sample with the magnetic imaging system. To image the irreversible laser ablation process, a specially designed sample holder containing numerous identical grids is mounted on a  two-dimension translation stage in xy plane.}
\end{figure*}

Proof-of-principle experiments of dynamic HEER were carried out at Tsinghua Thompson scattering X-ray source\cite{du2013generation} platfrom, which consists of a high-gradient normal-conducting RF gun working at 2.856 GHz, a 3-m long travelling wave accelerating tube and some other elements for beam control and diagnosis.  A schematic layout of this experiment is shown in Fig.\ref{fig:1}. High brightness electron probes with 300 pC bunch charge, 1 mm.mrad normalized emittance, 10 ps pulse duration, 45 MeV kinetic energy and 0.1$\%$ energy spread are generated from the photo-injector. At the exit of the accelerator, an electromagnetic quadrupoles composed triplet is used to focus the beam and match it into the imaging lenses section, which is entirely installed in the vacuum chamber. Electrons passing the the sample will travel through a PMQ based Russian quadruplet (RQ) and finally form a point-to-point image of the sample on a high-resolution YAG screen. Image of the sample will be recorded by a lens-coupled CCD outside of the chamber. A dipole magnet at the end of beamline is to calibrate the kinetic energy as well as measure the energy spread of the electron probe.

\begin{table}[htb]
\caption{\label{table1}%
Parameters describing the Russian quadruplet.
}
\begin{ruledtabular}
\begin{tabular}{lcdr}
\textrm{Name}&
\textrm{Length}&
\multicolumn{1}{c}{\textrm{Gradient}}&
\textrm{Position}\\
\colrule
$A_1$ & 18.63 mm & -186.1 $T/m$ & 0.02929 m \\
$B_1$ & 20.04 mm & 214.2 $T/m$  & 0.08366 m \\
$B_2$ & 20.16 mm & -209.0 $T/m$ & 0.13915 m \\
$A_2$ & 18.63 mm & 186.3 $T/m$  & 0.19348 m \\
\end{tabular}
\end{ruledtabular}
\end{table}

\begin{figure}[b]
\includegraphics[width=0.48\textwidth]{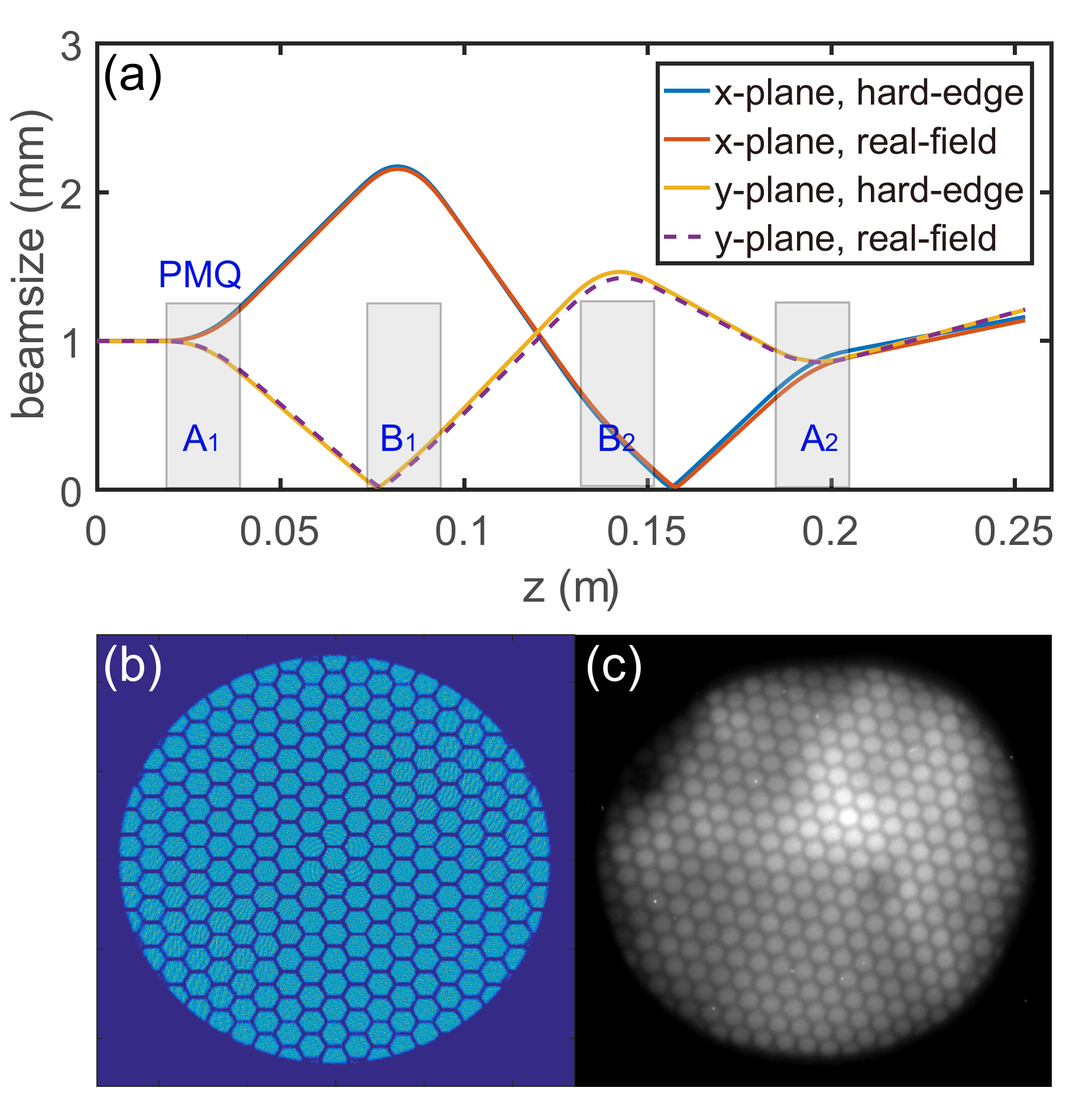}
\caption{\label{fig:2} (a): Simulated beam transverse envelop in imaging section. (b): simulated static image of a 200-mesh hexagon TEM grid. (c): experimental static image of a 200-mesh hexagon TEM grid.}
\end{figure}

\begin{figure}[htb]
\includegraphics[width=0.48\textwidth]{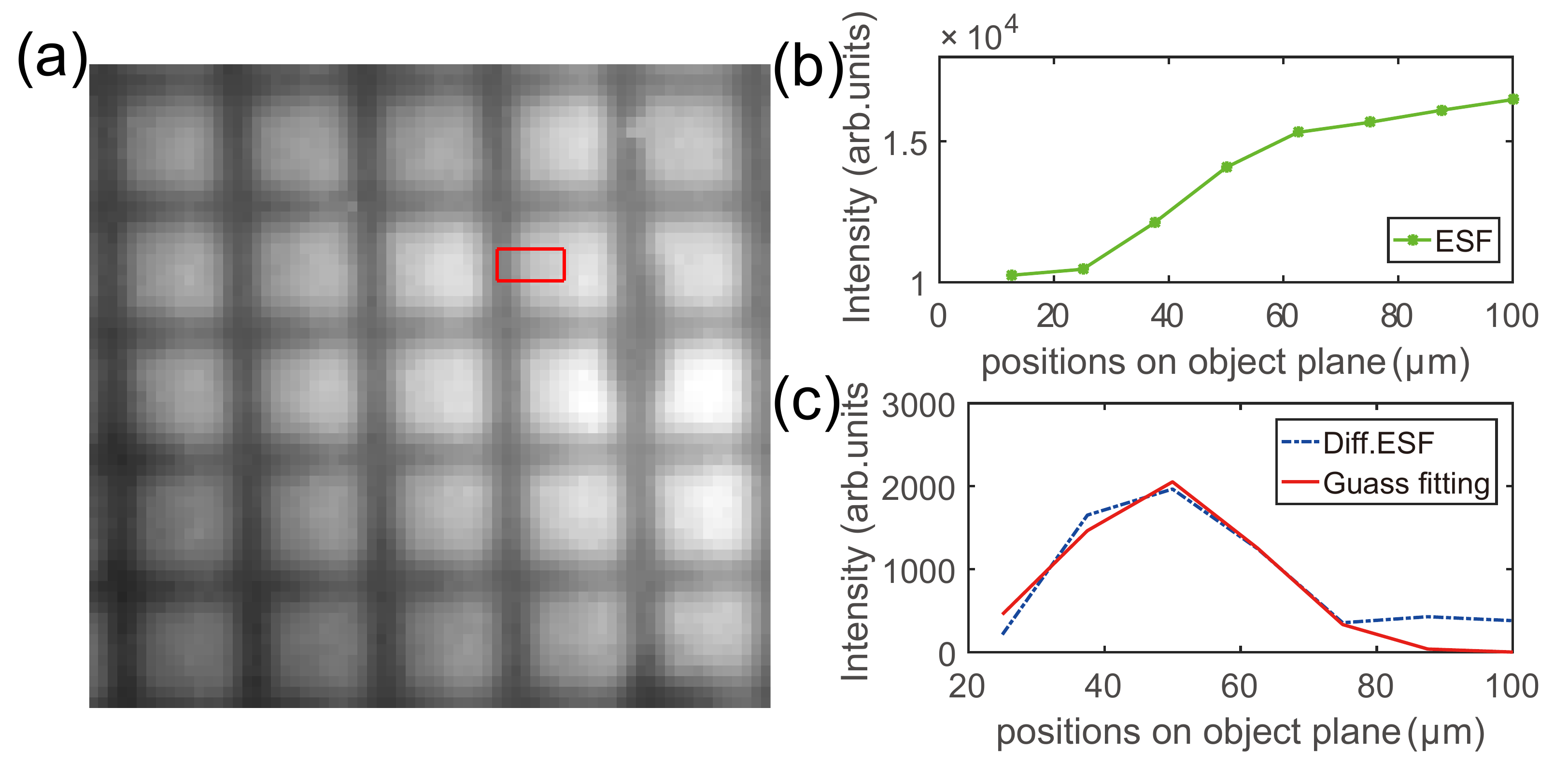}
\caption{\label{fig:3} (a): Image of the central part of a 200-mesh square TEM grid. (b): beam intensity distribution in x plane, i.e., edge spread function (ESF) of the grid bar in the region of interest marked by red box in (a). (c): the line spread function (differentiation of ESF) of the grid bar and its Gaussian fit plot. 1-sigma resolution is determined to be about 15 $\mu$m. }
\end{figure}

The RQ imaging lens is composed of two pairs of PMQs, which have found wide use in high energy particle transport\cite{schollmeier2008controlled}, focusing\cite{lim2005adjustable} and imaging\cite{li2014single,cesar2016demonstration}, due to their 1-2 order of magnitude higher gradients than common electromagnetic quadrupoles. Since their gradients cannot be tuned, positions of these PMQs are optimized by code COSY INFINITY \cite{makino2006cosy}. Parameters of the optimized layout of RQ are listed in Table. \ref{table1}. Fitness of the imaging section is verified by particle-tracking simulation using code ASTRA\cite{flottmann2013astra}, with both hard-edge PMQ model (same model used in COSY INFINITY code) and measured PMQ field distribution. In Fig. \ref{fig:2}, a point-to-point image of a 100$\%$ initial contrast sample (i.e, a standard 200 mesh hexagon TEM grid) is formed, with a magnification factor of about 1.1. The simulated imaging process matches well with experimental outcomes. Resolution of this imaging system is determined  to be 15 $\mu$m by measuring the edge spread function of a bar of the grid, as shown in Fig. \ref{fig:3}.

\subsection{Calibration of time-of-zero}
In pump-probe scheme, the first step for imaging the dynamic processes is to calibrate the time-of-zero, i.e, the beginning timestamp of laser initiated exciting of the sample. Electron deflectometry or shadowgraphy\cite{scoby2013single, peng2014time} is a widely used technique in determination of time-of-zero in laser-pump-electron-probe experiments, where electrons will be deflected by transient electromagnetic fields induced by high power density laser illumination of a metal target. In addition, this technique is found to be a promising tool for diagnosis of transient electric and magnetic fields\cite{chen2015mapping} and highly dense plasma evolving\cite{zhu2010four}. However, energy of electron probes used in previous experiments is limited to tens of keV up to a few MeV, since highly energetic electrons are more difficult to deflect. Advantage of higher energy electrons is also obvious, that is they could penetrate denser plasma and detect higher electric and magnetic fields\cite{chen2015mapping}.

\begin{figure}[htb]
\includegraphics[width=0.48\textwidth]{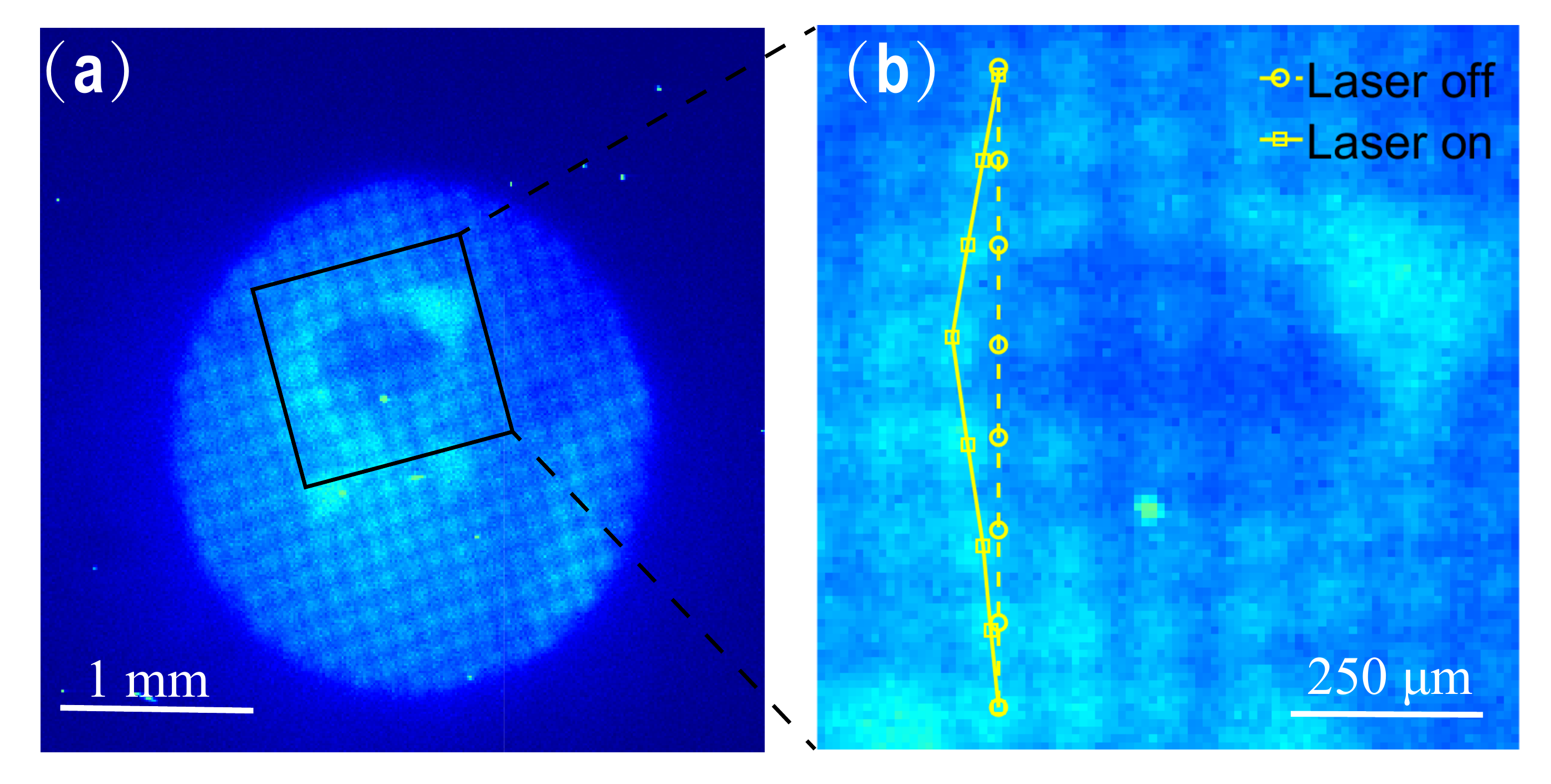}
\caption{\label{fig:4} (a): Electron deflection by transient electromagnetic fields at T=4.8 ns. (b): zoom view of the laser illuminating area.}
\end{figure}

In this experiment, an ultra-short high power intensity pump laser of 28 ($\pm 5\%$) mJ pulse energy, 40$\pm 2$ fs pulse duration (full width half height) is illuminating on a small area of $<$ 30 $\mu$m of the gold mesh, thus the power density is about 5$\times 10^{16}$ W/cm$^{2}$. With the RQ moved off the z axis, shadowgraphs of the laser irradiated sample are obtained. At a certain delay time, for example, T = 4.8 ns in Fig. \ref{fig:4}(a), beam intensity of the illuminating area drops while outer part intensity increases due to the deflection effect, thus forming a "valley-peak" intensity distribution pattern, which is very similar to those phenomena in previous experiment\cite{chen2015mapping}. Moreover, by measuring position offsets of centroid of each beamlet passing through the mesh holes with pump laser on, one can calculate the deflection angle of the electrons located in the "peak" area, which is about 1.7 mrad in maximum. By tuning the time delay between electron probe and pump laser, the lasting time of this phenomenon is determined to be approximate 8 ns.

\subsection{Imaging the melting process}
\begin{figure}[htb]
\includegraphics[width=0.48\textwidth]{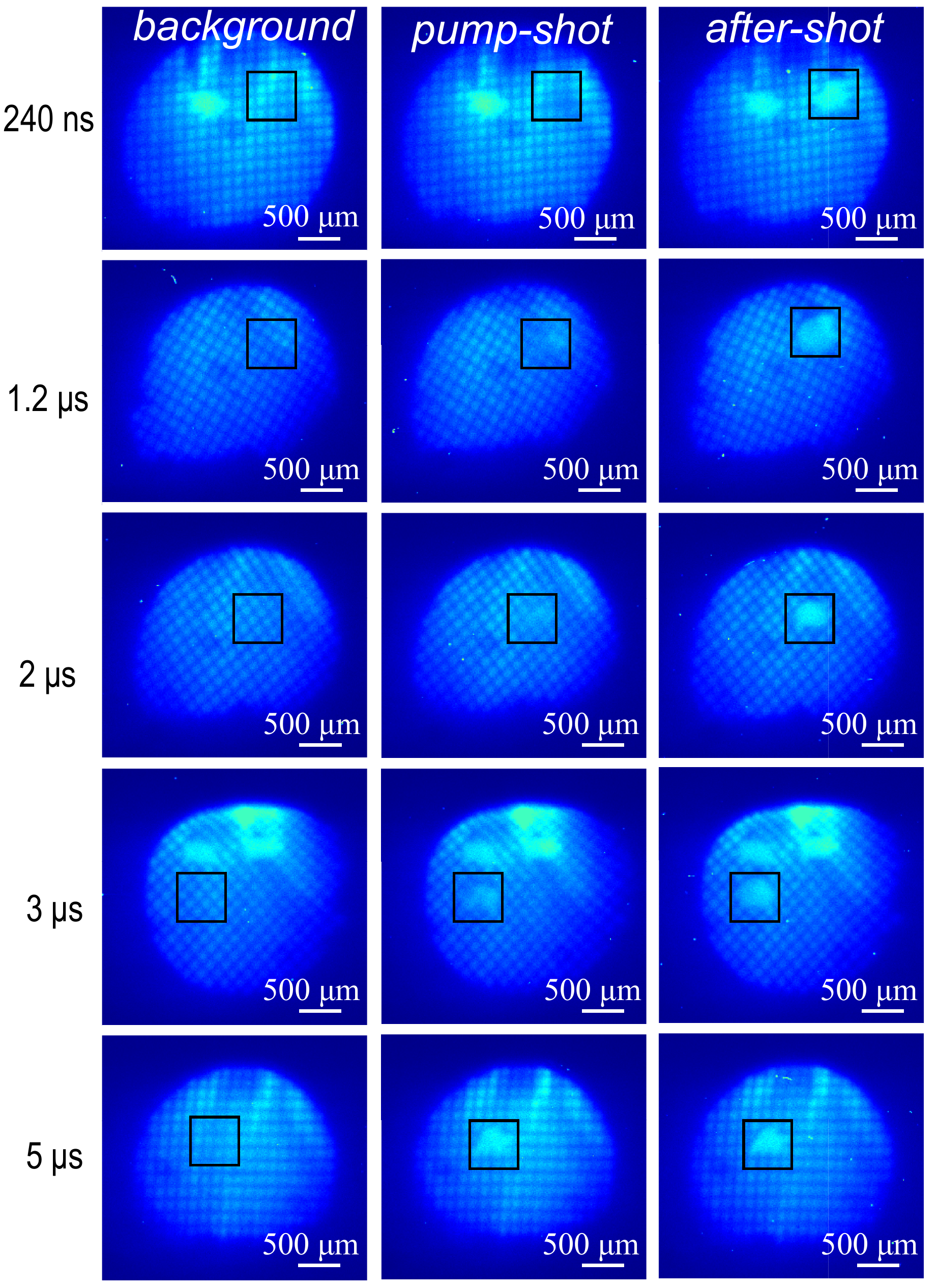}
\caption{\label{fig:5} Typical images of the gold mesh during the melting process. Black squares drawn in these images represent the illuminating area.}
\end{figure}

Known the time-of-zero, imaging mode was shifted from shadowgraph to radiograph mode, with RQ moved back to beam axis. Since the photo-injector is working on single pulse mode with a repetition rate of 10 Hz, the whole unrepeatable melting process can not be recorded with a single pulse. Therefore we designed a special sample holder containing numerous identical gold meshes, and by carefully varying time delay over a large dynamic scale, we can acquire a sequence of images of the sample at different time to piece together an entire movie of the melting process, as shown in Fig. \ref{fig:5}.

\begin{figure}[htb]
\includegraphics[width=0.5\textwidth]{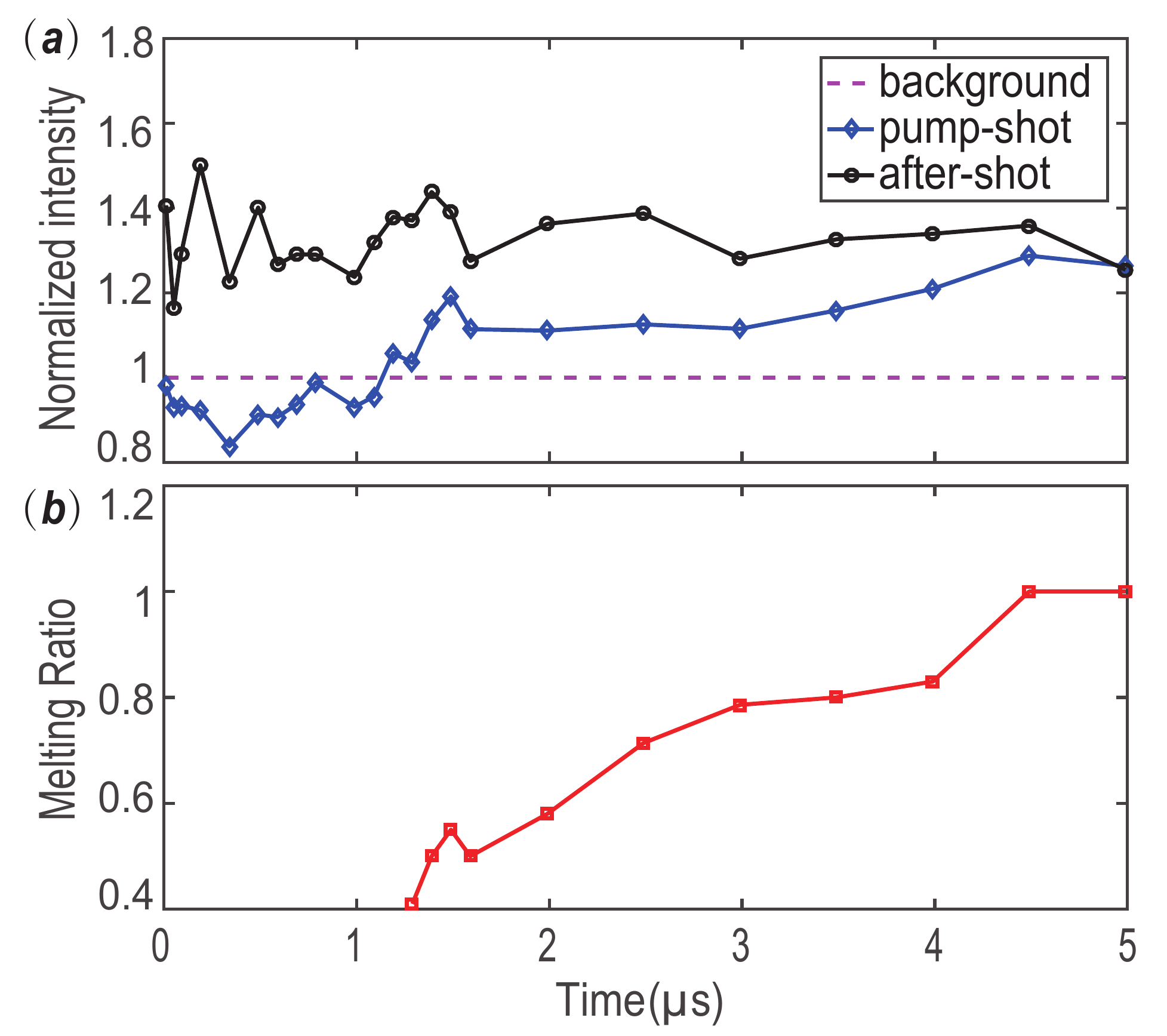}
\caption{\label{fig:6} Measured intensity of illuminating area (a) and melting ratio (b) at different time delays.}
\end{figure}

By analysing evolution of the beam intensity as well as the surface profile of the sample, the whole melting process can be divided into three different phases. In the first phase lasting from the beginning to about 10 ns, radiographs of the phase are almost same to the background-shots, unlike the obvious deflection observed in shadowgraph mode. This can be explained by the fact that the relatively small scatter angles imprinted on electrons by the gold mesh could not be effectively turned into image contrast in this point-to-point imaging mode. The second phase lasts from 10 ns to about 1.2 $\mu$s, featured by the intensity decreases while the surface profile of the sample remains almost unchanged, as shown in Fig. \ref{fig:6}. With the help of scanning electron microscope, surface split along the joint of adjacent pitches on the back of the grid can be observed, indicating that there is neutral-particle ejecta in the ablation process\cite{PhysRevLett.99.167602}. The last phase is from 1.2 $\mu$s to 5 $\mu$s, where the intensity of melting area increases gradually. In this phase, area with intensity higher than surrounding mesh holes is considered to be melted. By compare the area size in pump-shot radiograph and that in after-shot radiograph at each time delay, melting ratio is defined as $A_{pump-shot}/A_{after-shot}$. Changing of melting ratio versus time is illustrated in Fig. \ref{fig:6}(b), where melting ratio increases almost linearly until it approaches 1.

\section{SUMMARY}
In summary, dynamic high energy electron radiograph of laser-induced metal melting is demonstrated, with the combination of photo-injector generated high brightness electrons and a compact PMQ based imaging lens. With tens of ps-$\mu$m spatiotemporal resolution, dynamic high energy radiograph is proved to be suitable for high energy density matters and relevant fast dynamics processes. Moreover, spatial resolution of this dynamics HEER system can be further improved to $\mu$m level with a larger magnification factor imaging system, then more details of these dynamic processes can be observed.

\section{ACKNOWLEDGEMENT}
This work was supported by the National Natural Science Foundation of China (NSFC Grant No. 11435015 and No. 11505251).

\end{document}